# Fedora

## An Architecture for Complex Objects and their Relationships

Carl Lagoze, Sandy Payette, Edwin Shin, Chris Wilper

Computing and Information Science, Cornell University

{lagoze, payette, eddie, cwilper}@cs.cornell.edu

**Abstract.** The Fedora architecture is an extensible framework for the storage, management, and dissemination of complex objects and the relationships among them. Fedora accommodates the aggregation of local and distributed content into digital objects and the association of services with objects. This allows an object to have several accessible representations, some of them dynamically produced. The architecture includes a generic RDF-based relationship model that represents relationships among objects and their components. Queries against these relationships are supported by an RDF triple store. The architecture is implemented as a web service, with all aspects of the complex object architecture and related management functions exposed through REST and SOAP interfaces. The implementation is available as open-source software, providing the foundation for a variety of end-user applications for digital libraries, archives, institutional repositories, and learning object systems.

## 1 Introduction

As demonstrated by this special issue of *International Journal on Digital Libraries*, there is widespread interest in the representation, management, dissemination, and preservation of complex digital content. At a minimum, technologies for representing digital content should be able to match the richness, and complexity of well-established physical formats. As such, they should allow the representation of a variety of structural organizations, such as chapters and verses; accommodate the flexible combination of different genre of materials, such as text and images; and allow the aggregation of content from multiple sources and the association of metadata with the elements of the aggregation.

However, freed of the constraints of physical media, digital content architectures should do more. Exploiting their networked context, they should allow aggregation of content regardless of its physical location. By leveraging local and remote computing power they should support programmatic and user-directed manipulation of digital content. Finally, they should represent the complex structural, semantic, provenance, and administrative relationships among digital resources.

This paper describes our latest work on Fedora, an open-source digital content repository service, which provides a flexible foundation for managing and delivering complex digital objects. This recent work uniquely integrates advanced content management with semantic web technology. It supports the representation of *rich information networks*, where the *nodes* are complex digital objects combining data and metadata with web services and the *edges* are ontology-based relationships among these digital objects.

The motivation for integrating content management and the semantic web originates from requirements defined by the broader Fedora user community [5]. The most familiar example is the need to express well-known management relationships among digital resources such as the organization of items in a collection and structural relationships such as the part-whole relationships between individual articles and a journal. While the relationships among digital objects in these familiar applications are mainly hierarchical, we are working with other applications where the relationships are more graph-like. For example, in the NSF-funded NSDL (National Science Digital Library) Project [64], we are using Fedora to implement an *information network overlay* that represents local and distributed resources and the provenance, managerial, and semantic relationships among those resources. We report on the results of this work later in this paper.

While there are a number of schemes for representing these relationships such as conventional relational databases and formalisms like conceptual graphs [50], the products of the semantic web initiative such as RDFS [22], OWL [27], and highly-scalable triple-stores such as Kowari [55] provide extensible open-source solutions for representation, manipulation, and querying these knowledge networks.

The remainder of this paper describes the details of the Fedora architecture that provides the foundation for these rich applications. The structure of this paper is as follows. Section 2 briefly summarizes the historical development of Fedora. Section 3 provides core background on the Fedora digital object model, articulating a graph-based view of the model that is consistent with the semantic web orientation of our latest work. This is followed by Section 4 that describes the Fedora relationship model that provides a common framework for describing, storing, and querying relationships among objects and their components. In Section 5, we describe results of the deployment of Fedora, focusing on applications that exploit features that distinguish Fedora from related work. We conclude with a description of future work and appendices that provide a number of implementation details.

## 2 Background

The Fedora Project [6] is an ongoing research and development effort to provide the framework for creation, management, and preservation of existing and evolving forms of digital content. The roots of the project lie in DARPA-funded research in the early 1990's that defined the notion of a *digital object* [34] and implemented Dienst [36], a networked digital library architecture with protocol-based dissemina-tion of digital objects in multiple formats. Follow-on research extended these initial concepts with the notion of *active digital objects* [26] and *distributed active relationships* [25]. These concepts were refined and prototyped in a CORBA-based Fedora (Flexible Extensible Digital Object Repository Architecture) [42] as part of research with CNRI [41] and in the context of the NSF-funded Prism Project [2]. This prototype provided the context for a variety of research initiatives most notably in the areas of fine-grained policy enforcement [43] and preservation [44].

The transition of Fedora from a research prototype to production repository software began when the University of Virginia Library, seeking a solution for managing increasingly complex digital content, experimented with the Fedora architecture [51]. This experimentation took place in the context of innovations in humanities research [58]. The experimentation proved successful, providing the basis for subsequent funding from the Andrew W. Mellon Foundation to Cornell and Virginia [45] to jointly develop Fedora and make it available as open source software to libraries, museums, archives, and content managers, facing increasing variety and complexity in the digital content that they manage [52]. Mellon-funded development continues through 2007.

The richness of the Fedora digital object model and extensibility of the Fedora service-based architecture has led to its deployment in a variety of domains including digital libraries [35, 57], institutional repositories [53], electronic records archives [63], trusted repositories for digital preservation [33], library systems [62], educational technologies [56], web publishing [23], and distributed information networks [37].

Fedora is implemented as a set of web services that provide full programmatic management of digital objects as well and search and access to multiple representations of objects. All Fedora APIs are described using the Web Service Description Language (WSDL). As such, Fedora is particularly well-suited to exist in a broader web service framework and act as the foundation layer for a variety of multi-tiered systems, service-oriented architectures, and end-user applications. This distinguishes Fedora from other complex object systems that are turn-key, vertical applications for storing and manipulating complex objects through a fixed user interface (e.g., DSpace [49], arXiv [1], ePrints [3], Greenstone [10]).

By providing both a model for digital objects and repository services to manage them, Fedora is also distinguished from work focused on defining and promoting standard XML formats for representing and transmitting complex objects (e.g., METS [14], MPEG-21 DIDL[32], IEEE LOM [12]). However, Fedora is compatible with these efforts since it has the ability to ingest and export digital objects that are encoded in such XML transmission formats[1]. This allows Fedora to comfortably co-exist in the archival framework defined by OAIS [17].

---

[1] Fedora currently supports ingest/export of digital objects encoded using METS and also the Fedora XML wrapper format (FOXML). Future releases will support MPEG-21 DIDL and possible other formats.

As a service-based architecture for complex digital objects, Fedora has some commonality with the aDORe architecture [59] developed at the Los Alamos National Laboratory research library. The aDORe system provides a standards-based repository for managing and accessing complex digital objects. Objects are encoded in XML using DIDL [21] and a limited set of object relationships can be expressed using RDF. Object dissemination services are available via OAI-PMH [38] and OpenURL [40],

The current release of Fedora is version 2.0, which includes the semantic web integration that is a focus of this paper. Release 2.1 is scheduled for 3rd quarter 2005 and will include a powerful XACML-based [4] policy enforcement module described in a later section of this paper.

# 3 Fedora model for complex objects

The Fedora object model supports the expression of many kinds of complex objects, including documents, images, electronic books, multi-media learning objects, datasets, computer programs, and other compound information entities. Fedora supports aggregation of any combination of media types into complex objects, and allows the association of services with objects that produce dynamic or computed content. The Fedora model also allows the assertion of relationships among objects so that a set of related Fedora objects can represent the items in a managed collection, the components of a structural object like the chapters of a book, or a set of resources that share common characteristics (defined by semantic relationships).

Fedora defines a powerful object model for expressing this variety of complex content and their relationships. This object model can be understood from two perspectives.

1. The *representational* perspective defines a simplified abstraction for understanding Fedora objects, where each object is modeled as a uniquely identified resource projecting one or more views, or *representations*. From this perspective the internal structure of a digital object is opaque; however, relationships among objects are observable.
2. The *functional* perspective reveals the object components that underlie the representational perspective and provides the basis for understanding how the Fedora object model relates to the management services exposed in the Fedora repository architecture.

## 3.1 Representational View

The representational perspective of the Fedora object model asserts that each digital object can disseminate one or more representations of itself, and that each object can be related to one or more other objects. A familiar example of digital object with multiple representations is a document or image where the content is available in multiple formats. All digital objects, and their individual representations, are identi-fied with Uniform Resource Identifiers (URIs). These URIs are specified using the "info" scheme and conform to the syntax described at [13]. This choice frees the architecture from ties with any identifier resolution system (e.g., the Handle System [11].

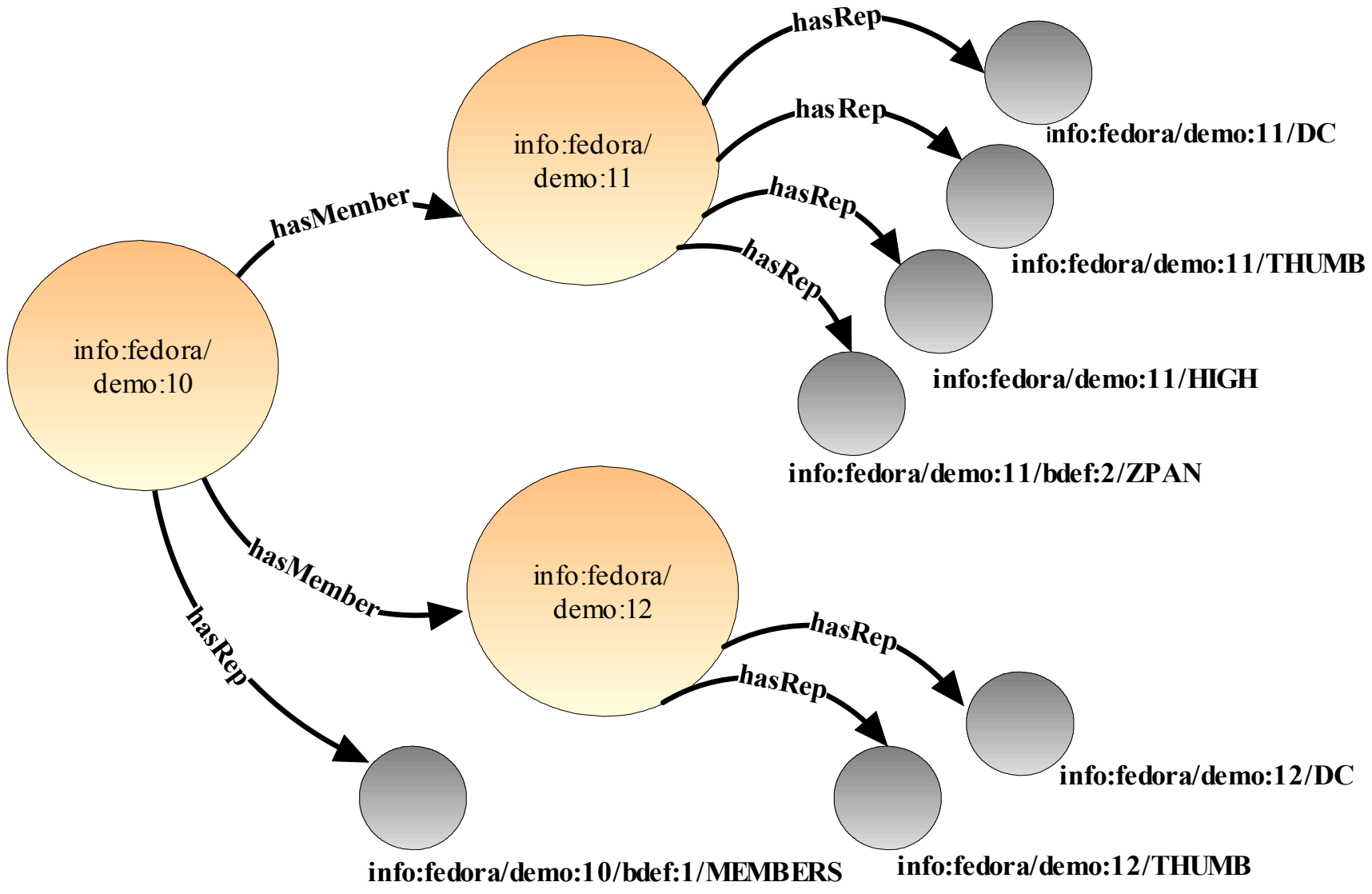

**Figure 1: Representational View of Fedora Objects**

This perspective hides complexity and exposes only the access points to content stored in a Fedora repository. Figure 1 depicts the representational view of three inter-related Fedora objects. The diagram shows a directed graph, where the larger nodes are digital objects, and the smaller nodes are representations of the digital objects[2]. These nodes are linked by two types of arcs – relationship arcs connect digital objects, and representation arcs connect digital objects to their respective representations. This graph can be expressed as RDF, stored in a triple store, and queried. This is discussed later Section 4.

Each digital object in the diagram has at least one representation, related to its originating digital object by a "hasRep" arc. For example, the node labeled info:fedora/demo:11 is an image digital object with four representations, identified by their respective URIs:

---

[2] This graph-based overlay model can form the basis for interoperability among heterogeneous object models and repositories. This concept is currently being explored as part of a new NSF-funded research project, Pathways, which is a collaboration between the authors of this paper and colleagues at Cornell, LANL, and others [16] [61].

- Dublin Core record, identified as info:fedora/demo:11/DC
- High-resolution image, identified as info:fedora/demo:11/HIGH
- Thumbnail image, identified as info:fedora/demo:11/THUMB
- Image with zoom/pan utility, as info:fedora/demo:11/bdef:2/ZPAN

We have yet to define the underlying source of these representations. In fact, in this view of the architecture such details are hidden from the client application concerned with access to these representations.

Figure 1 also demonstrates an example of inter-object relationships. In this example, the node labeled info:fedora/demo:10 is a "collection" with two "items", the nodes labeled info:fedora/demo:11 and info:fedora/demo:12. These collection-item relationships are expressed by the "hasMember" arc that emanates from the collection object. The inverse "isMemberOf" relationships are not shown in the diagram for simplification.

This simple representational view forms the basis of Fedora's REST-based access service (i.e., API-A-LITE), whereby digital object URIs and representation URIs can be easily converted to service request URLs upon Fedora repositories.

### 3.2 Functional View I - Datastreams

While the representational perspective of the Fedora object model provides a simple, access-oriented overlay for digital resources and collections, the *functional perspective* provides a view of the core underlying data model for Fedora. In the following sections, we take one of the digital object nodes depicted in Figure 1, and drill down to unveil the specific components of a Fedora digital object that enable access to representations. We start with the digital object as a container with a persistent unique identifier (i.e., PID). From there, we unveil the components incrementally, first focusing on components that enable simple content aggregation, then on components that enable dynamic and computed content, and finally on components related to digital object integrity. We note again that these underlying details are invisible to clients concerned only with information access.

In its simplest form a Fedora object is an aggregation of content items, where each content item maps to a representation. The Fedora object model defines a component known as a datastream to represent a content item. A datastream component either encapsulates bytestream content internally or references it externally. In either case that content may be in any media type. Figure 2 shows a digital object as an aggregation of datastreams and the one-to-one correspondence of those datastreams to the representations of the digital object that are exposed to accessing clients. In this simple case, each representation of a Fedora object is a simple transcription of the content that lies behind a datastream component.

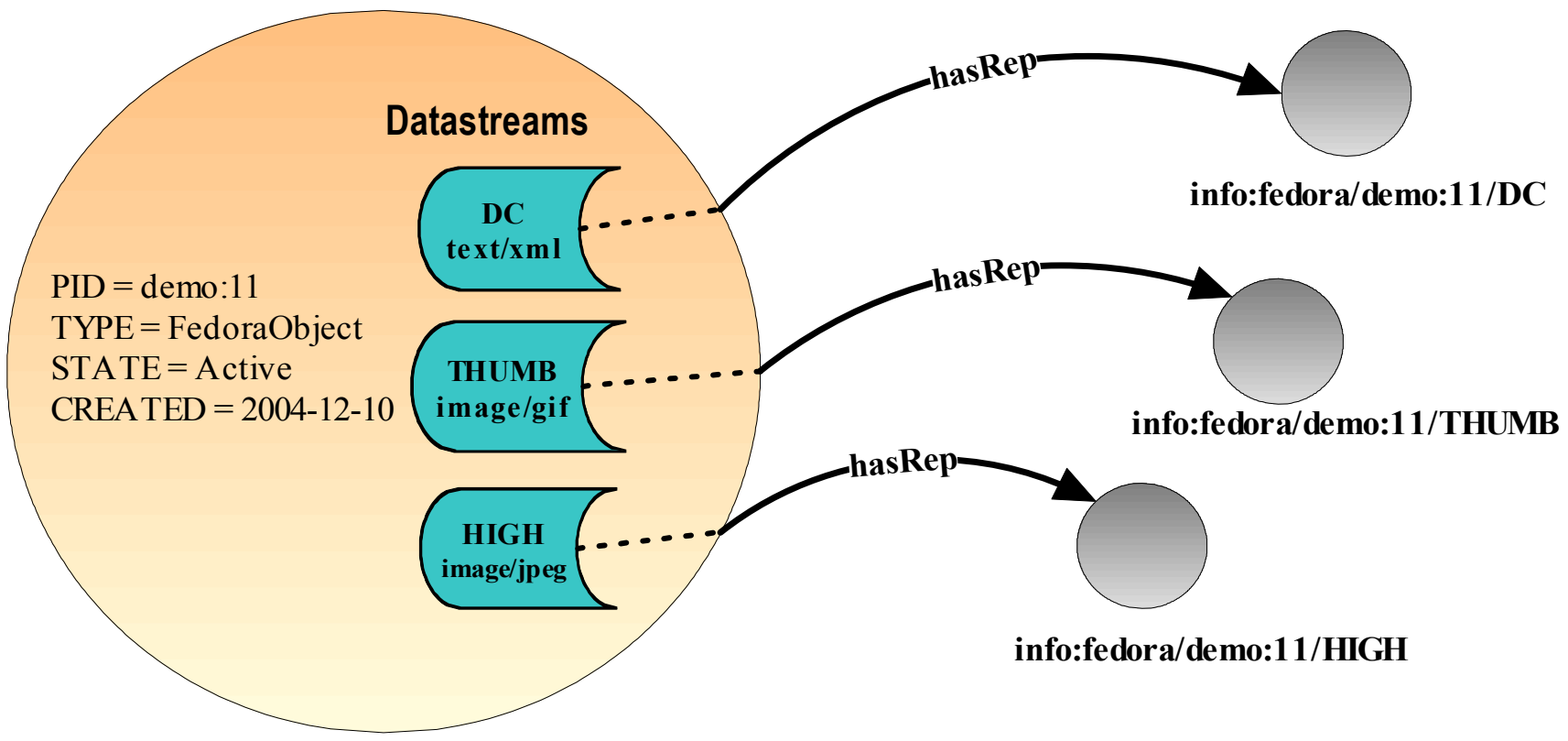

**Figure 2: Fedora Object with PID, Properties, and Datastreams**

As seen in the above diagram, a digital object has a unique identifier (PID) and a set of key descriptive properties. Each datastream contains information necessary to manage a content item in a Fedora repository. These are stored as properties of the datastream as shown in Figure 3.

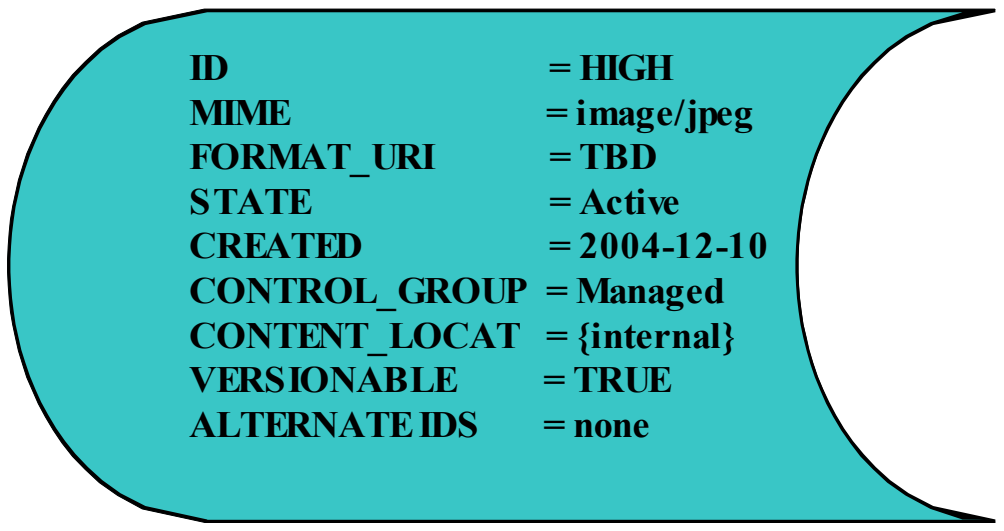

**Figure 3: Properties of a Datastream Component**

Three datastream properties deserve special attention. The Format URI refines the media type definition and anticipates the emergence of a global digital format registry such as the GDFR [9]. Control group defines whether the datastream represents either local or remote content. Datastreams with a control group of "Managed" are internal content bytestreams that are under the direct custodianship of the Fedora repository. Datastreams whose control group is "External" or "Redirected" (the difference between these is outside the scope of this paper) represent content that is stored outside the repository. These datastreams have a content location property that is a URL pointing to a service point outside the repository that is responsible for providing the content. The ability to create digital objects that aggregate locally managed content with external content is a powerful feature of Fedora, and is useful in a variety of contexts. A good example of a hybrid local/remote object is an educational object where local content is an instructor's syllabus, lecture notes, and exams, and remote content are primary resources included by-reference from other sites.

### 3.3 Functional View II - Disseminators

In addition to the representations described in the previous section, which are direct transcriptions of datastreams, the Fedora object model enables the definition of *virtual representations* of a digital object. A virtual representation, also known as a *dissemination*, is a view of an object that is produced by a service operation (i.e., a method invocation) that can take as input one or more of the datastreams of the respective digital object. As such, it is a means to deliver dynamic or computed content from a Fedora object.

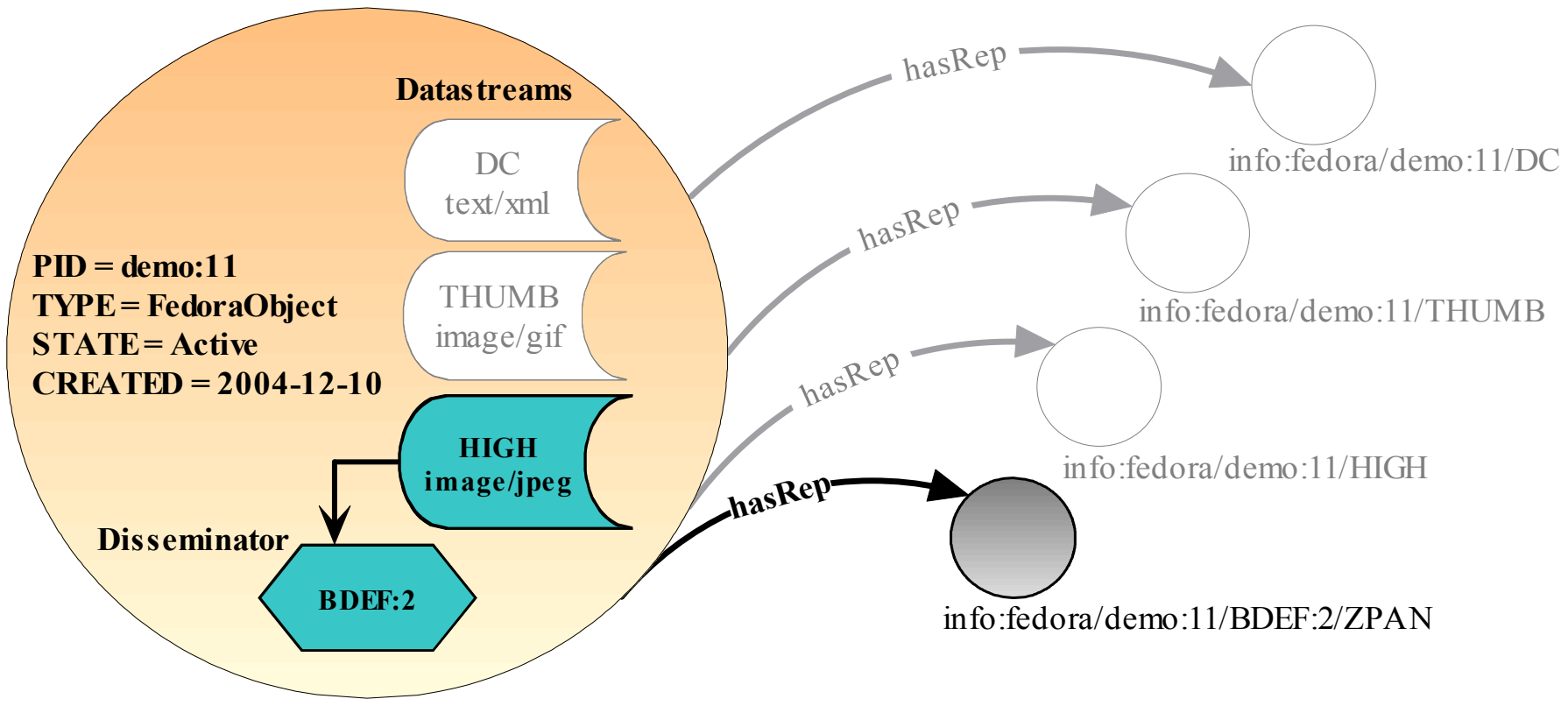

**Figure 4: Fedora Object with Disseminator Added**

This is illustrated in Figure 4, where a virtual representation labeled `info:fedora/demo:11/BDEF:2/ZPAN` is highlighted. From the access perspective this representation is an image wrapped in a java application that provides image zoom and pan functions. Note that this representation is not a direct transcription of any Datastream in the object. Instead, it is the result of a service operation defined in the Disseminator component labeled "BDEF:2" inside the object that uses the datastream labeled "HIGH" as input. The light-weight, REST-based interface to Fedora (API-A-LITE) makes it possible for a client application to pass parameters to the invoked service; in this case zoom and pan specifications.

To enable such behavior, a Disseminator must contain three pieces of information: (1) a reference to a description of service operation(s) in an abstract syntax, (2) a reference to a WSDL service description [19] that defines bindings to concrete web service to run operation(s), and (3) the identifiers of any Datastreams in the object that should be used as input to the service operation(s).

Fedora stores the service operation description and the WSDL service description within special digital objects, respectively known as BDefs (behavior definitions) and BMechs (behavior mechanisms). Figure 5 depicts a Fedora BDef object and a BMech object along with object-to-object relationships that exist due to the presence of the Disseminator component in the main object (i.e., demo:11).

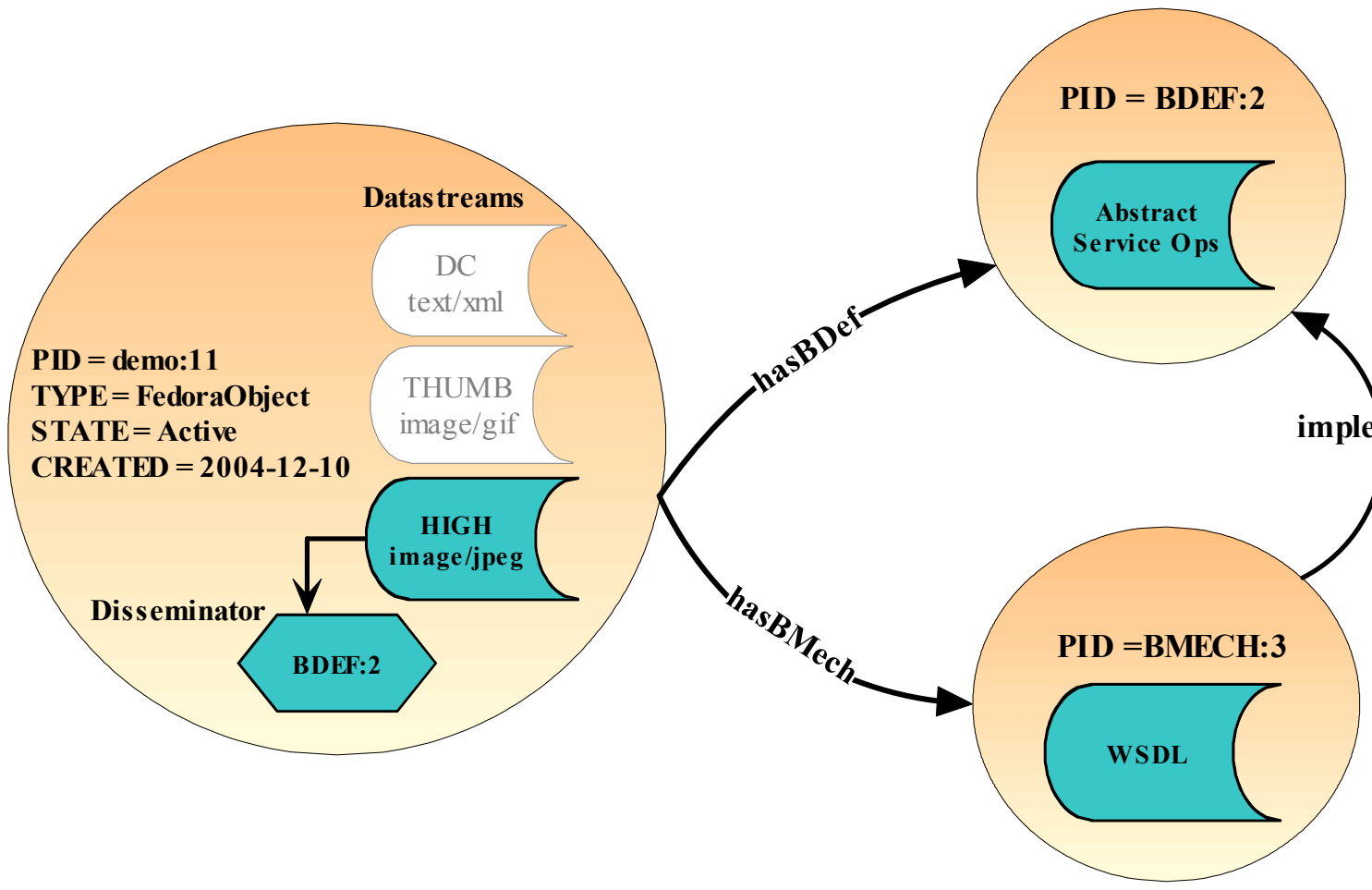

**Figure 5: Disseminators establish relationships to service definition objects**

Disseminators are effectively metadata that a Fedora repository uses at run time to construct and dispatch service requests and produce one or more virtual representations of the digital object. From a client perspective this is transparent since virtual representations look just like other representations of the object.

Disseminators are a powerful feature in the Fedora object model. They can be used to create common representational access points for digital objects that have different underlying structure or format. For example, an institutional repository might contain scholarly documents in a variety of root formats (e.g., Word, HTML, TeX), where the root format is stored as a datastream in a Fedora digital object. For interoperability purposes, a virtual representation can be defined on each object that converts the datastream containing the root format to a common format (e.g., PDF). Similarly, a repository manager can decide for archival purposes to convert all documents in a repository to a canonical preservation format without disrupting the manner in which clients access documents for browsing, viewing, etc. Finally, disseminators can add utility operations to digital objects. For example, a Disseminator can be defined for a digital object that provides parameterized query access to the relationships defined for that object. Such a query might return the "members of a collection" or, in the case of an educational digital library such as the NSDL [64], the set of resources that are appropriate for K-12 mathematics education. The implementation of these queries is described in Section 4.

### 3.4 Functional View III – Object Integrity Components

The Fedora object model defines several metadata entities that pertain to managing the integrity of digital objects. These entities are the object's relationship metadata, access control policy, and audit trail. To keep the Fedora model simple and consistent, integrity entities are modeled as datastream components with reserved identifiers. As such, the integrity entities are stored like other datastreams, however the Fedora Repository system recognizes them as special and asserts constraints over how they are created and modified. Figure 6 depicts these integrity-oriented entities as special datastreams in a digital object, identified as Relations, Policy, and Audit Trail.

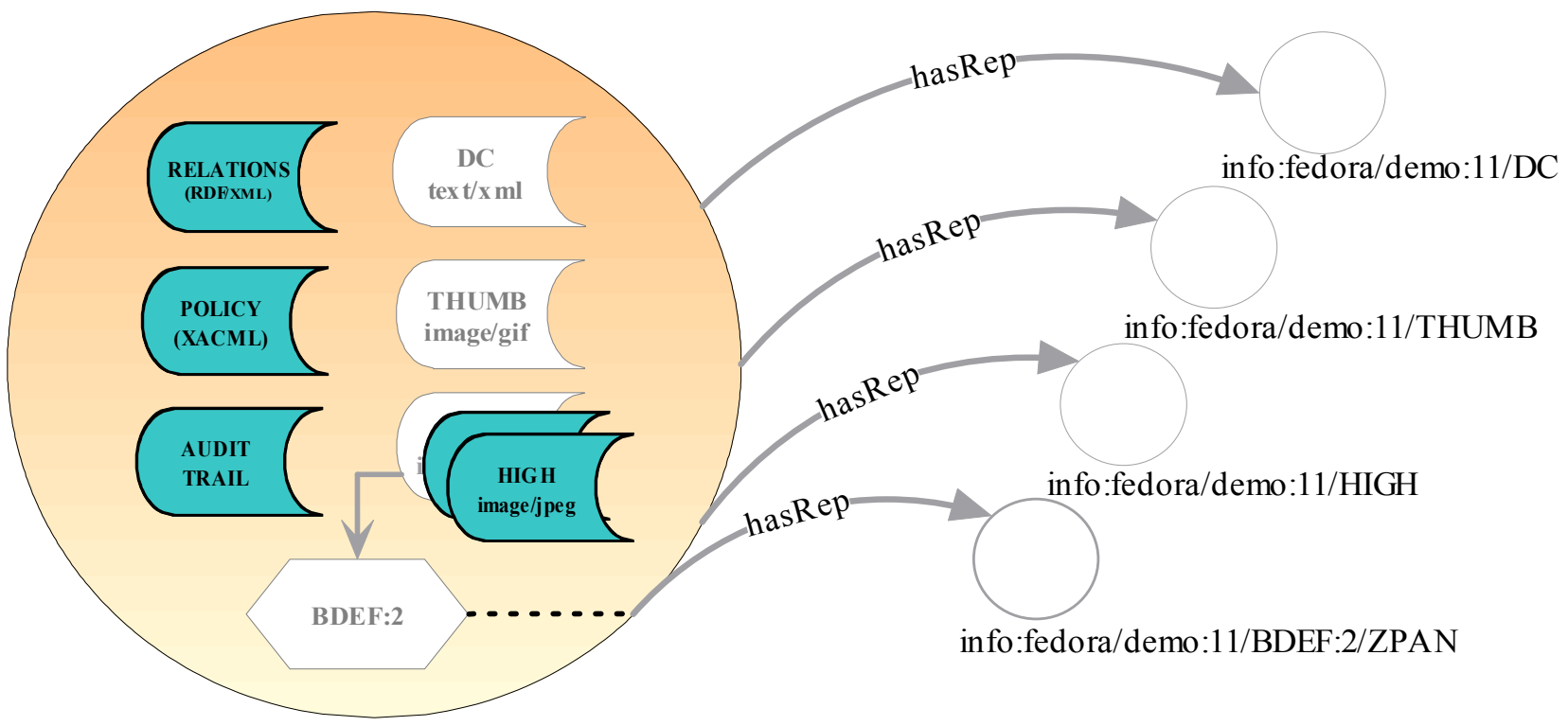

**Figure 6: Integrity Datastreams - Relationships, Policy, and Audit Trail**

A Relations datastream is used to assert object-to-object relationships such as collection/member, part/whole, equivalence, "aboutness," and more. The previously discussed "hasMember" relationship is an example of the type of assertion that can be managed via the Relations datastream, described in Section 4.

A Policy datastream is used to express authorization policies for digital objects, both to protect the integrity of an object and to enable fine-grained access controls on an object's content. In Fedora objects, a policy is expressed using the eXtensible Access Control Markup Language (XACML) [4], which is a flexible XML-based language used to assert statements about who can do what with an object, and when they can do it. Object policies are enforced by the authorization module (i.e., AuthZ) implemented within the Fedora repository service.

The Audit Trail is a system-controlled datastream that keeps a record of all changes to an object during its lifetime. The Fedora repository service automatically creates an audit record for every operation upon an object, detailing who, what, when, where, and why an object was changed. This information is important to support preservation and archiving of digital objects.

Another feature for managing the lifecycle of objects is versioning. Versioning is important for applications where change tracking is essential, as well as for preservation and archiving systems that must be able to recover historical views of digital objects. The Fedora object model supports component-level versioning, meaning that datastreams and disseminators can be changed without losing their former instantiations. Fedora automatically creates a new version of these components whenever they are modified.

This is depicted in Figure 6, which shows a digital object with multiple versions of a datastream (see component labeled "HIGH"). Also, the versioned datastream is input to the disseminator labeled "BDEF:2." Requests for representations of this digital object can be date-time stamped and the Fedora repository service will ensure that the appropriate component version is returned. This feature applies for representations that are direct transcriptions of datastream content, as well as for virtual representation where datastream content is mediated via a Disseminator.

## 4 Relationships in Fedora

As described in Section 2.1, the Fedora object model can be abstractly viewed as a directed graph, consisting of *internal* arcs that relate digital objects nodes to their representation nodes and *external* arcs between digital objects. In this section we focus on that relationship graph and describe the Fedora *Resource Index* module, which allows storage and query of the graph. This module builds on RDF (Resource Description Framework) [39] primitives developed within the semantic web community. The Fedora system defines a base relationship ontology that, in the fashion of any RDF properties, can co-exist with domain-specific ontologies from other namespaces. Each digital object's relationships to other digital objects are expressed in RDF/XML [20] within a reserved datastream in the object. The Resource Index is a relationship graph over all digital objects in the repository that is derived by merging the relationships implied by the Fedora object model itself with the relationships explicitly stated in an object's relationship datastream. The triples representing this graph are then stored in a Kowari [55] triple-store providing the capability for searching over the graph.

The combination of representing explicit relationships as RDF/XML in a datastream of a digital object and then mapping them to the Kowari triple store offers the "best of both worlds". The explicit representation provides the basis for exporting, transporting, and archiving of the digital objects with their asserted relationships to other objects. The mapping to Kowari provides a graph-based index of an entire repository and the basis for high-performance queries over the relationships[3]. An added advantage of the dual representation is that the entire triple store can be rebuilt by importing and parsing the XML-based digital objects.

---

[3] Our preliminary report on query performance of various triple store technologies is available at http://tripletest.sourceforge.net/2005-06-08/index.html. We are planning a future publication that explores these results and their implications.

### 4.1 Representing object-to-object relationships

There has been a significant amount of work in the area of structural metadata for complex objects. These efforts have been focused on developing XML schema for expressing structural relationships with individual digital objects. One early example was the Making of America [28] project that formalized structural metadata and defined a set of templates that correspond to well-known physical artifacts such as a book composed of chapters and diaries consisting of entries. The current exemplar of this encoding of structural metadata is in METS [15].

Our focus in Fedora has been to decompose these structural units into separate digital objects. The motivation for this is that the units can then be reused in a variety of structural compositions. In addition, this lays the basis for expressing other types of non-structural relationships among digital objects such as:

- The organization of individual resources into larger *collection* units, for the purpose of management, OAI-PMH harvesting [38], user browsing, and other uses.
- The relationships among *bibliographic* entities such as those described in the Functional Requirements for Bibliographic Relationships [8].
- *Semantic* relationships among resources such as their relevance to state educational standards or curricula in an educational digital library like the National Science Digital Library [64].
- Modeling more complex forms of *network overlays* over the resources in a content repository such as citation links [29, 31], link structure, friend of a friend [7], etc.

All of these relationships, including structural relationships, should be expressible both within individual digital objects and among multiple digital objects. For example breaking the components of a structural entity, such as the chapters of a book, into separate digital objects provides the flexibility for reuse of those individual components into other structural units. This is even more important for the other forms of relationships. For example, a single resource may be part of multiple collections or may be relevant for multiple state standards.

The remainder of this section describes a relatively simple example of inter-object relationships to demonstrate how these are expressed in Fedora. The simple techniques illustrated here can be used to express more complex inter-object relationships. In Section 5, we will describe a more complex example in the context of our NSDL work.

The expression of arbitrary, inter-object relationships in Fedora is enabled by a reserved datastream known as the Relations datastream. This datastream allows for a restricted subset of RDF/XML where the subject of each statement must be the digital object within which the datastream is defined.

```
<rdf:RDF
   xmlns:rdf ="http://www.w3.org/1999/02/22-rdf-syntax-ns#"
   xmlns:nsdl=”http://nsdl.org/std#”
   xmlns:rel="http://example.org/rel#"
   xmlns:frbr="http://example.org/frbr#">
     <rdf:Description rdf:about="info:fedora/demo:11">
       <rel:isMemberOf  rdf:resource="info:fedora/demo:10"/>
       <std:fulfillsStandard rdf:resource="info:fedora/demo:Standard5"/>
       <frbr:isManifestionOf rdf:resource=
             "info:fedora/demo:Expression2"/>
    </rdf:Description>
</rdf:RDF>
```

**Table 1 - Example Relations datastream**

Since predicates from any vocabulary can be used in Relations, the repository manager has considerable flexibility in the kinds of relationships that can be asserted. Table 1 shows an example Relations datastream in a Fedora digital object identified by the URI, info:fedora/demo:11. The RDF/XML refers to three different relationship vocabularies (hypothetical for the purpose of this example) and asserts the following relationships:

- demo:11 is a member of the collection represented by the object demo:10,
- demo:11 fulfills the state educational standard represented by the object demo:Standard5,
- demo:11 is a manifestation of the expression represented by the object demo:Expression2.

### 4.2 Object representations and properties in the Resource Index

As described earlier, a Fedora digital object consists of a number of core components such as datastreams and disseminators, which bind to BDefs and BMechs. In addition each Fedora digital object has system metadata or properties. The architecture provides a system-defined ontology to represent the relationships among these core components. For example, the relationships of an object to its representations is expressed using the <fedora-model:disseminates> predicate as shown in the triple in Table 2.

```
<info:fedora/demo:11>
  <fedora-model:disseminates>
     <info:fedora/demo:11/HIGH>
```

**Table 2 - Object-represenation relationship**

In addition to these relationships, the system-defined ontology also represents object data properties whose range contains date and boolean datatypes, as shown in the triple in Table 3.

```
<info:fedora/demo:11/HIGH>
   <fedora-view:lastModifiedDate>
      "2004-12-12T00:22:00"^^xsd:dateTime
```

**Table 3 - Data type properties**

Unlike the relationships expressed in the Relations datastream, these relationships are not explicitly asserted within the digital object. Instead they are derived from the structure of the object itself and mapped into the Resource Index, alongside the relationships represented in the Relations datastreams. This is described in the next section.

### 4.3 Storing and querying the relationship graph

All these relationships – the relationships explicitly stated in the Relations datastream, the relationships implied by the object structure, and the data relationships contained in the object properties – are stored in the Resource Index. This index is automatically updated by the repository service whenever an object structure is modified or its Relations datastream is changed.

The Resource Index handles queries over these relationships. The combination of all relationships into a single graph, and the automated management of that combined graph, enables a powerful and flexible service model. External services may issue queries combining relationships from different name spaces, since they are all RDF properties. For example, Table 4 shows a query listing all the representations of all objects that are members of a particular collection.

```
select $dissemination
from <#ri>
where ($object <fedora-view:disseminates> $dissemination)
    and $object <rel:isMemberOf> <demo:10>
```

**Table 4 - Sample RDF query using iTQL**

An early design goal of the Resource Index was to allow the use of different triplestores and thus permit the Fedora repository administrator to choose the most appropriate underlying store. To that end, the Resource Index employs a triplestore API similar in spirit to JDBC, to provide a consistent update and query interface to a variety of triplestores. Extensive testing of both query performance time and query language features ultimately led to the selection of Kowari as the default triplestore[4].

The query interface to the relationship graph currently supports three RDF query languages, RDQL [48], iTQL [54], and SPO [47][5]. Both RDQL and iTQL share a superficially similar syntax to SQL, with RDQL enjoying broader implementation support, but iTQL providing a richer feature set [30].

The RDF query results naturally take the form of rows of key-value pairs, again similar to the result sets returned by a SQL query. However, it is often useful to work with a sub-graph or a constructed graph based on the original. To this end, the query API may also return *triples* instead of *tuples*.

### 4.4 Using the relationship graph

The Resource Index is exposed as one of the interfaces of the core Fedora repository service. This facilitates the development of other services in the Fedora Service Framework that is described in Appendix A. The Resource Index interface is exposed in a REST architectural style to provide a stateless query interface that accepts queries by value or by reference. The service has been implemented with an eye toward eventual conformance to the W3C Data Access Working Group's SPARQL protocol for RDF[24], as it matures.

One example of a service exploiting the Resource Index is the OAI Provider Service that exposes metadata about resources in a Fedora repository. This OAI Provider Service is quite flexible in that it can be configured to allow harvesting not only of static metadata formats, but those that are dynamically produced via service-based disseminations of Fedora objects.

```
select $member $collection $dissemination
from <#ri>
where $member <rel:isMemberOf> <info:fedora/demo:10>
    and $member <rel:isMemberOf> $collection
    and $member <rel:isMemberOf> $dissemination
    and $member <fedora-view:disseminates> $dissemination
    and $dissemination <fedora-view:disseminationType>
<info:fedora/*/bdef:OAI/getQualifiedDC>
```

**Table 5 - A query to build an OAI response**

An example of the interaction of this service with the Resource Index is as follows. An external OAI harvester requests qualified Dublin Core records for a particular set of resources from the repository. The OAI Provider service processes this by issuing the query to the Resource Index listed in Table 5. This query effectively requests "all *disseminations* of qualified Dublin Core records of resources that are members of the collection identified as 'demo:10'". The significance of requesting disseminations is that the Dublin Core records may not statically exist as datastreams within the object, but they may be derived from another metadata format such as MARC. The Resource Index query would return the tuples shown in Table 6 that can provide the basis of an OAI response. Note that the OAI representations were not shown earlier in Figure 1.

[4] Our preliminary report on query performance of various triple store technologies is available at http://tripletest.sourceforge.net/2005-06-08/index.html. We are planning a future publication that explores these results and their implications.

[5] Future releases will also support SPARQL [46].

| member | collection | dissemination |
|---|---|---|
| info:fedora/ demo:11 | info:fedora/ demo:10 | info:fedora/ demo:11/ bdef:OAI/getDC |
| info:fedora/ demo:12 | info:fedora/ demo:10 | info:fedora/ demo:12/ bdef:OAI/getDC |

**Table 6 - The query response as tuples**

## 5 Results

Fedora has been tested in the field with the real-world collections of our collaborators. These applications demonstrate the flexibility of the Fedora object model and repository service to accommodate a diverse set of information management problems. They distinguish Fedora from seemingly similar architectures such as DSpace, arXiv, and ePrints, whose focus is primarily on institutional repositories for scholarly publications. The applications supported by Fedora include not only complex digital library collections [35, 57] and institutional document repositories [53], but also electronic records archives [63], trusted repositories for digital preservation [33], and distributed information networks such as the NSDL. This section describes results of Fedora deployment in three of these contexts: the University of Virginia digital library collections; the Encyclopedia of Chicago, a multimedia cultural heritage resource; and the National Science Digital Library, a distributed information network of educational resources and contextual information about those resources.

The core functionality of Fedora has proven effective for integrating rich digital collections at University of Virginia Library (UVA). The UVA digital repository is built upon well-defined "content models" for digital content, where a content model specifies the number and types of datastreams and disseminators for particular genre of complex digital objects, including images, books, letters, archival finding aids, and data sets [57]. The result is a seamless integration of content in a repository that enables consistent management of digital objects, consistent interfaces to access digital objects, and easy re-use of digital materials in different contexts. The architecture provides a means to easily aggregate materials from different collections and create new views on content using both Fedora relationship metadata and custom disseminators. One example is a cross-collection object that builds upon multiple objects: one that disseminates architectural drawings about historical buildings, another that contains recent photographs of those buildings from art collections, and another that contains historical letters that mention the buildings from the electronic text collections.

Northwestern University's use of Fedora demonstrates how Fedora's flexibility allows the management and publication of rich multimedia objects. Most compelling is the electronic version of the Encyclopedia of Chicago [23] produced in collaboration with the Chicago Historical Society. The encyclopedia is a multi-media resource that exploits the Fedora disseminator capability in novel ways. A notable feature of this application is the design of digital objects and disseminators to create rich, clickable maps. These maps are linked to disseminators that provide multi-dimensional views and contextual information about a location in Chicago. For example a street map of Chicago highlights sites of labor unrest. By clicking on the map, a user can discover and launch numerous disseminations that link to population statistics, newspaper articles, and historical data that relate to a particular place on the map. This is all done using well-designed digital object content models and rich, service-based disseminators to produce dynamic transformations of digital object content. Nearly every piece of content on the web site is a dissemination of a Fedora digital object, and interestingly, the entire web site itself is published via a single dissemination of a master collection object.

Our work in the context of the NSF-funded NSDL (National Science Digital Library) Project [64] is perhaps the most interesting example of the power of Fedora's relationship architecture. Our goal in the NSDL is not only to provide a digital library allowing search and access to distributed resources, but to augment NSDL resources with context that defines their usability and reusability in different learning and teaching environments. By "context", we mean information such as the provenance of the resources, the manner in which resources have been used, comments by users that annotate and explain primary resources, and linkages between the resources and relevant state educational standards. While the NSDL work is specifically targeted at the education domain, we argue that the notion of *contextualization* is increasingly important as a means of adding value to digital content and defining its quality based on provenance, utility, and other factors.

Using the content management and semantic web tools in Fedora we have implemented an *information network overlay* [37]. This architecture represents the data underlying the NSDL as a graph of typed nodes, corresponding to the information entities in the NSDL, and semantic edges representing the contextual relationships among those entities. The nature and variety of these relationships will evolve over time and, thus, any fixed schema approach for representing the network overlay would be too restrictive. Our results thus far indicate that the semantic web approach of Fedora is particularly well-suited for this application.

Figure 7 illustrates a fragment of the network overlay. The nodes in the overlay graph correspond to Fedora digital objects – each shape corresponding to an information entity in the NSDL. These entities include agents, resources, metadata, and the like. The edges are relationships among these entities, which are represented in the Resource Index. For example, Figure 7 shows the grouping of resources in collections, and the provenance trail of who originally recommended those resources and who manages them. Relationships from other ontologies, such as state education standards, are overlaid on this base graph. These are similarly represented in the Resource Index alongside the base ontology relationships. The entire knowledge base can then be queried by external services to build rich portals for users and tools for inferring quality, usability, and educational value.

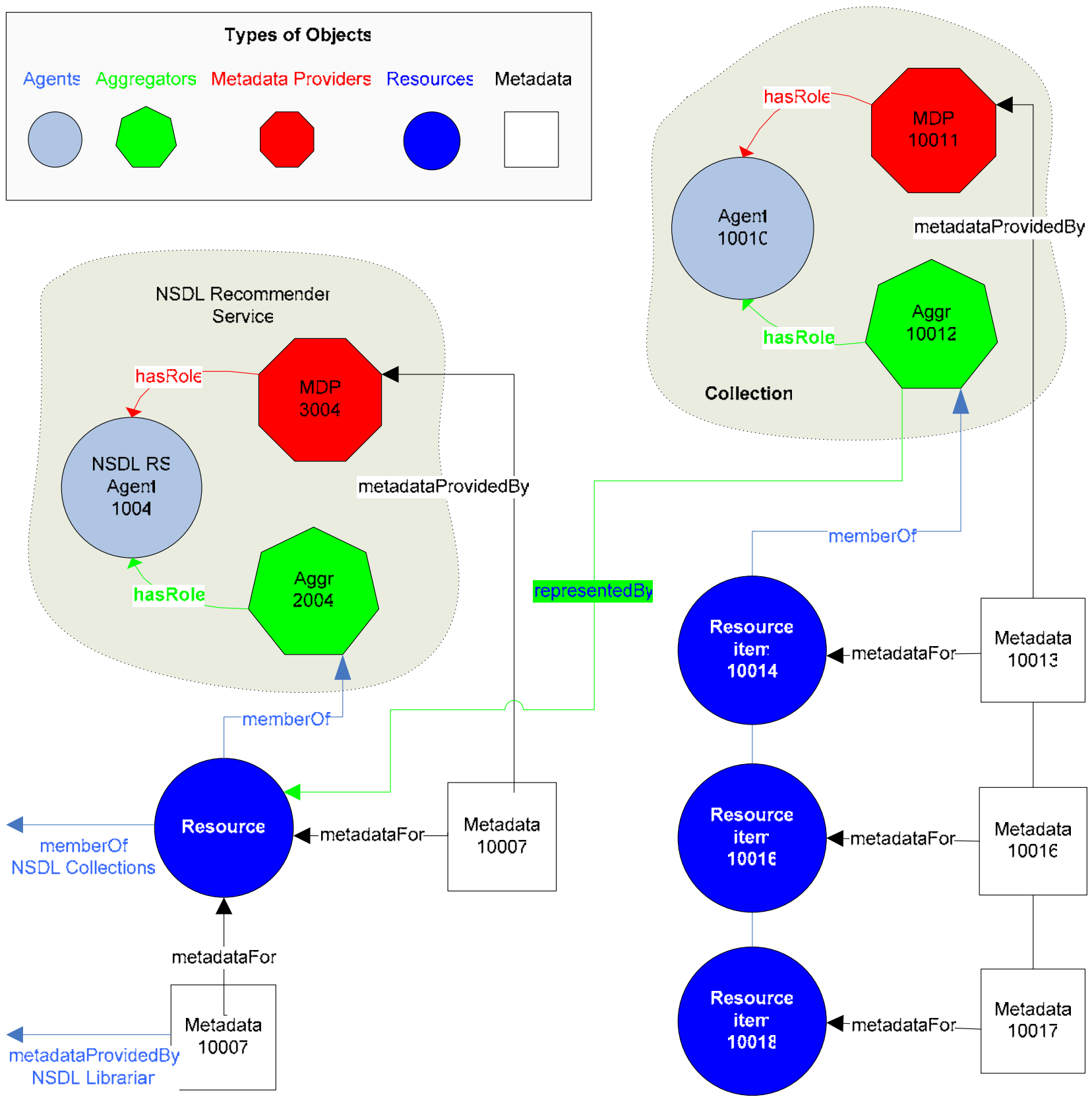

**Figure 7 - NSDL network overlay example**

## 6 Conclusion and Future Work

As mentioned previously, the Fedora project is now in its second phase of funding from the Andrew W. Mellon Foundation. This phase extends through 2007. This work addresses a number of areas including adding new services, tools, and utilities; optimizing for scale and performance; adding new integrity and preservation features; and enabling the creation of peer-to-peer networks of repositories (federations). This work will be motivated by the specific requirements of institutional repositories, extremely large digital collections and archives, and distributed educational applications. The phase 2 development plan is prioritized to first focus on functionality that will make it easier for institutions to get a jump start in using Fedora – specifically by easily loading heterogeneous digital collections into Fedora repositories. Later work we will add functionality that helps Fedora users move towards building large-scale, highly dependable repositories. This will provide the basis for a shared, seamless information space in which virtual collections and networked objects can be fully realized.

In addition to this core development work by Cornell and the University of Virginia teams, a number of other parties have joined in a Fedora Development Consortium[6]. The purpose of this group is to provide a framework for collective knowledge sharing and collaboration for developers working within the Fedora Service Framework, described earlier. The Consortium will augment the core Fedora system with additional value-added open-source software, and eventually produce a number of vertical "Fedora-inside" applications. For example, the NSDL team is collaborating with the Fedora project to implement a content management system over Fedora that includes workflow and a configurable user interface. Also, several working groups have been commissioned to develop services for preservation, Shibboleth-based [18] authentication, and more powerful search mechanisms.

Fedora has been designed from the beginning for extensibility. A key aspect of its basic design is the existence of a well-defined object model and the exposure of the model through programmatic interfaces. A powerful feature of this model is the notion of an object having multiple representations, including virtual representations that involve the interaction of data and services. Another important feature of the model is the extensible relationship architecture that allows content managers to model within Fedora complex networks of information. Finally, the Fedora Service Framework, which is the implementation context for this object model, is the foundation for the deployment of extended services and user/client applications that apply Fedora in a variety of domains.

Increasingly rich digital content is placing greater demands on the institutions responsible for the creation, storage, management, and preservation of that content.

---

[6] Details on members of the consortium are at the Fedora Open Source Project web site – http://www.fedora.info.

Fedora is well-positioned to meet those demands and its open architecture provides the basis for meeting new requirements as they develop in the future.

## Acknowledgments

The Fedora Open Source Project is currently funded by the Andrew W. Mellon Foundation. The authors express gratitude to Mellon for their long-term support of this project. Prior support came from the National Science Foundation and DARPA. The authors are indebted to the Fedora team members at the University of Virginia: Thornton Staples (co-PI) and Ross Wayland (technical lead), Ronda Grizzle, Bob Haschart, Bill Niebel, and Tim Sigmon. The Fedora project has also benefited from the collaboration and contributions of members of the Fedora Development Consortium, especially Bill Parod at Northwestern University. The authors also thank NSDL team members Tim Cornwell and Elly Cramer. NSDL work is funded by the National Science Foundation under grant numbers 0127308 and 0127520. The NSF is not responsible for the content of this paper. Finally, the authors express gratitude to Michael Nelson and Herbert Van de Sompel for their on-going collaboration and their encouragement in writing this paper, and to the reviewers of the initial draft of this paper for their helpful comments.

## Appendix A The Fedora Repository Service

The digital object model described in this paper exists within the context of a broader server architecture. The appendix describes that architecture. Further details are documented at the Fedora open-source project web site [6].

### A.1 The Fedora Service Framework

Fedora digital objects are managed within the Fedora Service Framework which consists of a set of loosely coupled services that interact and collaborate with each other. At the core of the framework is the Fedora repository service, as depicted in Figure 8. Other services exist around the core to provide additional functionality that is not considered a fundamental function of a repository. Any number of services can be developed to collaborate with the core Fedora repository service. In the diagram, there are three collaborating services around the core: the Fedora OAI provider, a Fedora Search service, and a Fedora Preservation Monitoring Service. The framework approach anticipates that new services will be added over time.

Outside of the boundaries of the Fedora framework are external services that can either call upon Fedora services, or that Fedora can leverage in some way. The distinction between services within the Fedora Service Framework, and those outside, is that those within the framework are in a trusted relationship with the Fedora repository service, and are designed to specifically interact with Fedora repositories. Services outside the framework are typically general-purpose services, or organization-specific services that call upon Fedora as an underlying repository for digital content.

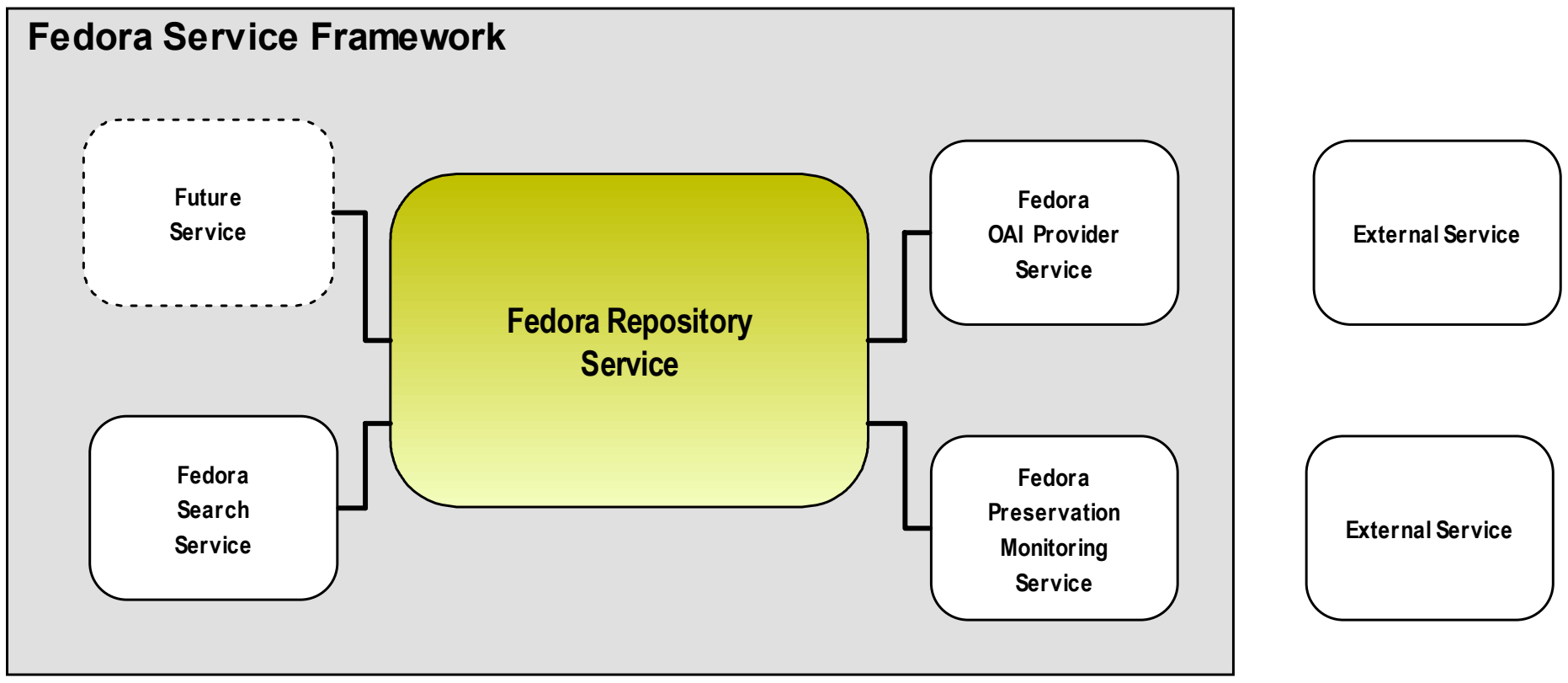

**Figure 8: Core Fedora Repository Service with Collaborating Services**

Prior to version 2.0 of Fedora, all Fedora-related functionality was built into the core Fedora repository service. As of version 2.0, the Fedora Service Framework was defined to move the Fedora architecture in a direction where new services can easily be developed and plugged into the Framework. This is consistent with general trends developing in web services technology and enterprise application architectures in which formerly tightly-integrated systems are broken apart into atomic, modular services that can be flexibly aggregated into different multi-service compositions.

At the time of writing, Fedora is migrating to the new service framework approach. Version 2.1 of Fedora will include a new OAI Provider and a new Search service as part of the Fedora open-source distribution. These functions were previously built into the core repository. The Fedora Preservation Monitoring Service will be developed as part of the new Phase II Fedora project. Other services are being developed by members of the Fedora user community and will be contributed back to the open source project.

### A.2 The Fedora Repository Service

At the core of the Fedora Service Framework is the Fedora repository service which exposes interfaces for managing and accessing digital objects in a repository. In Figure 9, the repository service is deconstructed so that its internal modules and public service interfaces are visible.

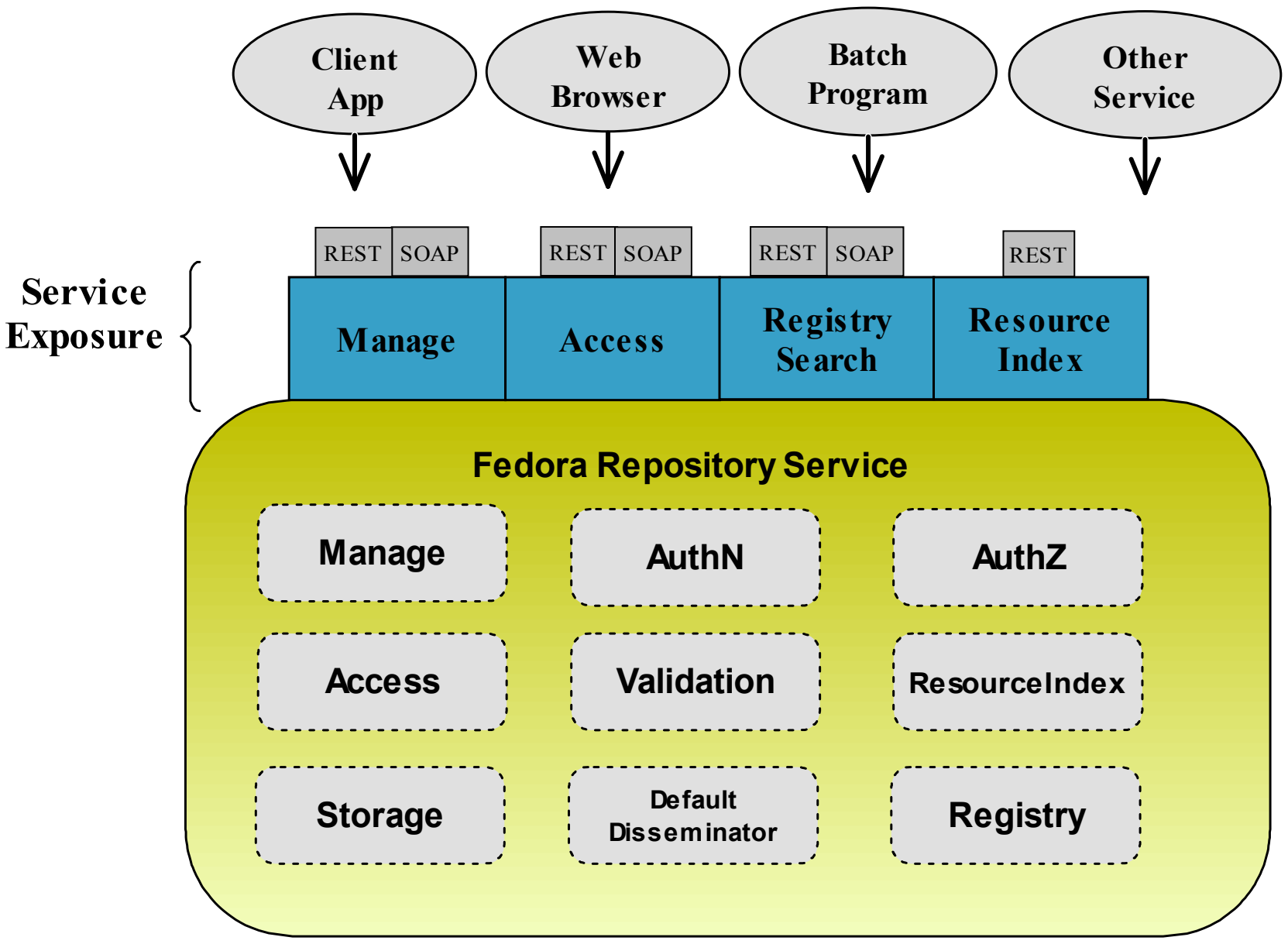

**Figure 9: Fedora Repository Internal Modules and Service Interfaces**

At the top of Figure 9 there are alternative client scenarios for accessing the Fedora repository through its four web service interfaces. Each service interface is defined using the Web Service Description Language (WSDL) [19], with both SOAP and REST bindings. The internal implementation of the Fedora repository service consists of a set of internal java modules that can be configured, and optionally replaced with alternative implementations. The internal modules are not directly exposed to accessing clients; instead clients interact with the repository only through the defined web service interfaces.

The Manage(ment) service interface (API-M) contains read/write operations necessary to manage a repository of digital objects. API-M operations exist for ingesting and exporting digital objects in an XML format, either Fedora's FOXML, or alternatively METS or MPEG21/DIDL. Also, objects can be created and modified using component-level operations that reflect the functional view of the Fedora object model described earlier. The major management operations are:

- set/get/removeObjectProperty
- set/get/removePolicy
- add/modify/purgeDatastream
- add/modify/purgeDisseminator
- ingest/export/purgeObject

The Access service interface (API-A) contains read-only operations for accessing digital objects. The two main purpose of the Access interface is to (1) introspect on a digital object (i.e., to discover what datastreams and disseminator methods are available) and (2) request disseminations on an object (i.e., access particular representations of the object's content). The major Access operations are:

- getObjectProfile
- getObjectHistory
- listDatastreams
- listMethods
- getDatastreamDissemination
- getDissemination

In addition to the SOAP-based Access bindings, all Access operations can be invoked with a simple URL syntax via a light-weight, REST-based interface (API-A-LITE). This interface can be used to access digital objects from the representational perspective described earlier. The graph node URIs for Fedora objects and their representations can be easily converted to Fedora API-A-LITE request URLs by replacing the "info:fedora" URI scheme with the base URL for the repository as follows:

```
info:fedora/demo:11        →  http://myfedora.edu:8080/fedora/get/demo:11
```

```
info:fedora/demo:11/HIGH → http://repo.edu:8080/fedora/get/demo:11/HIGH
```

The final two access points to the Fedora repository service are the Registry Search and Resource Index interfaces. These provide discovery capabilities to locate digital objects. The Registry Search interface exposes service operations to perform a simple search of the digital object registry based on object properties. The Resource Index interface is the service entry point to an RDF-based index of the entire repository. The Resource Index is an expanded version of the representational view of digital objects described earlier. As such it contains all representations and relationships of objects, plus object properties and Dublin Core metadata elements.

## Appendix B XML Serialization of Fedora Objects

This paper has presented both a representational and functional perspective of the Fedora object model. These provide an understanding of the abstractions that form the basis of a Fedora digital object. From an implementation perspective, Fedora digital objects can be serialized and stored as XML. The Fedora object model is directly expressed using XML Schema language in a format known as Fedora Object XML (FOXML)[7]. FOXML defines a <digitalObject> root element that contains a set of <objectProperties>, one or more <datastream> components, and one or more <disseminator> components. .

Although FOXML is the preferred XML serialization format for storing objects in a Fedora repository, Fedora supports ingest and export of digital objects in other XML formats. Currently, the system supports a Fedora profile of the Metadata Encoding and Transmission Format (METS) [15] and it will soon support the OAI-PMH harvesting [60] of objects encoded in MPEG21 Digital Object Description Language (DIDL) [32].

The XML serialization of the digital object info:fedora/demo:11 that corresponds to the example described earlier in this paper is as follows.

```
<?xml version="1.0" encoding="UTF-8"?>
<foxml:digitalObject PID="demo:11"
  xmlns:foxml="info:fedora/fedora-system:def/foxml#"
  xmlns:xsi="http://www.w3.org/2001/XMLSchema-instance"
  xsi:schemaLocation="info:fedora/fedora-system:def/foxml#
  http://www.fedora.info/definitions/1/0/foxml1-0.xsd">
    <!-- ********************************************************************** -->
    <!-- OBJECT PROPERTIES -->
    <!-- ********************************************************************** -->
    <foxml:objectProperties>
      <foxml:property NAME="http://www.w3.org/1999/02/22-rdf-syntax-ns#type"
                      VALUE="FedoraObject"/>
      <foxml:property NAME="info:fedora/fedora-system:def/model#state"
                      VALUE="A"/>
      <foxml:property NAME="info:fedora/fedora-system:def/model#label"
                      VALUE="Image Object – UVA Pavilion"/>
      <foxml:property NAME="info:fedora/fedora-system:def/model#createdDate"
                      VALUE="2004-12-10T00:21:57Z"/>
      <foxml:property NAME="info:fedora/fedora-system:def/view#lastModifiedDate"
                      VALUE="2004-12-23T00:20:00Z"/>
      <foxml:property NAME="info:fedora/fedora-system:def/model#contentModel"
                      VALUE="UVA_STD_IMG"/>
    </foxml:objectProperties>
```

[7] The FOXML schema is available at http://www.fedora.info/definitions/1/0/foxml1-0.xsd.

```xml
<!-- ********************************************************************** -->
<!-- DATASTREAMS -->
<!-- ********************************************************************** -->
<foxml:datastream ID="THUMB" CONTROL_GROUP="E" MIMETYPE="image/jpg"
                    STATE="A" VERSIONABLE="true">
     <foxml:datastreamVersion ID="THUMB.0" LABEL="Preview Pavilion III"
                                CREATED="2004-12-10T00:21:57Z">
          <foxml:contentLocation TYPE="URL"
               REF="http://icarus.lib.virginia.edu/images/iva/archerd05small.jpg" />
          </foxml:datastreamVersion>
</foxml:datastream>
<foxml:datastream ID="HIGH" CONTROL_GROUP="M" MIMETYPE="image/jpeg"
                    STATE="A" VERSIONABLE="true">
     <foxml:datastreamVersion ID="HIGH.0" LABEL="Drawing Pavilion III"
                                CREATED="2004-12-10T00:21:57Z">
          <foxml:contentLocation TYPE="INTERNAL_ID"
                                 REF="demo:11:HIGH:HIGH.0"/>
     </foxml:datastreamVersion>
     <foxml:datastreamVersion ID="HIGH.1" LABEL="Drawing Pavilion III"
                                CREATED="2004-12-12T00:22:00Z">
          <foxml:contentLocation TYPE="INTERNAL_ID"
                                 REF="demo:11:HIGH:HIGH.1"/>
     </foxml:datastreamVersion>
     <foxml:datastreamVersion ID="HIGH.2" LABEL="Drawing Pavilion III"
                                CREATED="2004-12-23T00:20:00Z">
          <foxml:contentLocation TYPE="INTERNAL_ID"
                                 REF="demo:11:HIGH:HIGH.2"/>
     </foxml:datastreamVersion>
</foxml:datastream>
<!-- ********************************************************************** -->
<!-- INTEGRITY DATASTREAMS -->
<!-- ********************************************************************** -->
<foxml:datastream ID="DC" CONTROL_GROUP="X"  MIMETYPE="text/xml"
                    STATE="A" VERSIONABLE="true">
     <foxml:datastreamVersion ID="DC.0" LABEL="Dublin Core Record"
                                 CREATED="2004-12-10T00:21:57Z">
          <foxml:xmlContent>
          <oai_dc:dc xmlns:oai_dc="http://www.openarchives.org/OAI/2.0/oai_dc/"
         xmlns:dc="http://purl.org/dc/elements/1.1/">
                    <dc:title>Image of UVA Pavilion - Drawing</dc:title>
                    <dc:subject>Architectural drawings</dc:subject>
                    <dc:publisher>University of Virginia</dc:publisher>
                    <dc:identifier>demo:11</dc:identifier>
          </oai_dc:dc>
          </foxml:xmlContent>
     </foxml:datastreamVersion>
</foxml:datastream>
<foxml:datastream ID="RELS-EXT" CONTROL_GROUP="X" MIMETYPE="text/xml"
                    STATE="A" VERSIONABLE="true">
          <foxml:datastreamVersion ID="RELS-EXT.0" LABEL="Relationships"
                                     CREATED="2004-12-10T00:21:57Z">
               <foxml:xmlContent>
                    <rdf:RDF xmlns:rdf="http://www.w3.org/1999/02/22-rdf-syntax-ns#"
                 xmlns:rdfs="http://www.w3.org/2000/01/rdf-schema#"
```

```
                xmlns:fedora="info:fedora/fedora-system:def/relations-external#"
                xmlns:myns="http://www.nsdl.org/ontologies/relationships#"
                xmlns:dc="http://purl.org/dc/elements/1.1/"
                xmlns:oai_dc="http://www.openarchives.org/OAI/2.0/oai_dc/">
                    <rdf:Description rdf:about="info:fedora/demo:11">
                        <fedora:isMemberOf rdf:resource="info:fedora/demo:10"/>
                    </rdf:Description>
                </rdf:RDF>
            </foxml:xmlContent>
        </foxml:datastreamVersion>
    </foxml:datastream>
    <foxml:datastream ID="AUDIT" CONTROL_GROUP="M" MIMETYPE="text/xml"
                STATE="A" VERSIONABLE="false">
        <foxml:datastreamVersion ID="AUDIT.0" LABEL="Object Audit Trail"
                        CREATED="2004-12-12T00:22:00Z">
            <foxml:xmlContent>
                <audit:auditTrail xmlns:audit="info:fedora/def:audit/">
                    <audit:record ID="AUDREC1">
                        <audit:process type="Fedora API-M"/>
                        <audit:action>modifyDatastreamByRef</audit:action>
                        <audit:componentID>HIGH</audit:componentID>
                        <audit:responsibility>fedoraAdmin</audit:responsibility>
                        <audit:date>2004-12-12T00:22:00Z </audit:date>
                        <audit:justification></audit:justification>
                    </audit:record>
                    <audit:record ID="AUDREC2">
                        <audit:process type="Fedora API-M"/>
                        <audit:action>modifyDatastreamByRef</audit:action>
                        <audit:componentID>HIGH</audit:componentID>
                        <audit:responsibility>fedoraAdmin</audit:responsibility>
                        <audit:date>2004-12-23T00:20:00Z</audit:date>
                        <audit:justification></audit:justification>
                    </audit:record>
                </audit:auditTrail>
            </foxml:xmlContent>
        </foxml:datastreamVersion>
    </foxml:datastream>
    <!-- ********************************************************************** -->
    <!-- DISSEMINATOR(S) -->
    <!-- ********************************************************************** -->
    <foxml:disseminator ID="DISS1" BDEF_CONTRACT_PID="BDEF:2" STATE="A"
                VERSIONABLE="true">
        <foxml:disseminatorVersion ID="DISS1.0"
                BMECH_SERVICE_PID="BMECH:3"
                LABEL="UVA Simple Image Behaviors"
                CREATED="2004-12-10T00:21:57Z">>
            <foxml:serviceInputMap>
                <foxml:datastreamBinding KEY="HIGHRES_IMG"
                    DATASTREAM_ID="HIGH" LABEL="Input Image"/>
            </foxml:serviceInputMap>
        </foxml:disseminatorVersion>
    </foxml:disseminator>
</foxml:digitalObject>
```